\documentclass[revsymb,aps,prb,12pt]{revtex4}

\begin{document}
\title{The variational symmetries and conservation laws in classical theory
of Heisenberg (anti)ferromagnet.}
\author{I.G. Bostrem}
\author{A.S. Ovchinnikov}
\author{R.F. Egorov}
\address{Department of Theoretical Physics, Ural State University, \\
620083, Lenin Ave. 51, Ekaterinburg, Russia}

\date{\today}
\begin{abstract}
The nonlinear partial differential equations describing the spin
dynamics of Heisenberg ferro and antiferromagnet are studied by
Lie transformation group method. The generators of the admitted
variational Lie symmetry groups are derived and conservation laws
for the conserved currents are found via Noether's theorem.
\end{abstract}


\maketitle

\newpage {}

\section{Introduction.}

As it has been demonstrated previously in a number of works the symmetry
methods are very efficient tool in application to differential equations in
physics. A subject of an especial interest is a study of invariance
properties (symmetries) of the equations with respect to local Lie groups
point transformations of dependent and independent variables. The detailed
discussion and applications of the symmetries of differential equations and
variational problems may be found in Ref.\cite{ibr,Olver} and \cite
{ovs}.

The non-linear dynamics of $n$-dimensional (anti)ferromagnet is described by
non-linear second-order partial differential equations (PDE). In our
previous work it has been shown that the system of differential equations
are the Euler-Lagrange equations of a certain action functional and the
corresponding long-wave action has been constructed via Vainberg's theorem
(see also Appendix). It is well known that a part of one parameter symmetry
groups of these equations turns out to be their variational symmetries as
well, i.e. a symmetries of the associated action functional. According to
Noether theorem \cite{ibr} such an invariance of the elementary action is a
necessary and sufficient condition of an existence of conservation laws for
the smooth solutions of the initial PDE.

The layout of the paper is as follows. By using the Noether identity we
define the variational symmetries among the point Lie symmetries of the
initial differential equations of a $2D$-dimensional ferromagnet. After they
are deduced we establish explicit formulae for the conserved densities and
density currents involved in the conservation laws both for a Heisenberg
ferromagnet and antiferromagnet. Apparently, the analysis presented may be
easily generalized for an arbitrary dimension $n\geq 3$.

\section{Dynamical invariants of $2D$ ferromagnet.}

The non-linear partial differential equations in $3$ independent variables $%
\left( x^{1},x^{2},x^{3}\right) =(t$, $\vec{r})$ and two dependent variables
$u^{1}=\theta (t,\vec{r})$ and $u^{2}=\varphi (t,\vec{r})$ describing the
dynamics of Heisenberg ferromagnet in continuum approximation can be written
as follows \cite{bishop}
\[
\mathrm{F}^{1}\equiv -S\sin \theta \varphi _{t}+\beta \left( \Delta \theta -\cos
\theta \sin \theta \left( \vec{\nabla}\varphi \right) ^{2}\right) =0,
\]
\begin{equation}
\mathrm{F}^{2}\equiv S\sin \theta \theta _{t}+\beta \left( 2\cos \theta \sin \theta
\left( \vec{\nabla}\theta \vec{\nabla}\varphi \right) +\sin ^{2}\theta
\Delta \varphi \right) =0,  \label{eq:ferroeq}
\end{equation}
where $\beta =\frac{JS^{2}}{\hbar }$. Here and throughout the following
notations are taken: $\alpha $ have the range $1$, $2$ meaning $\theta $ and
$\varphi $, correspondingly, the usual summation convention over a repeated
index is employed, $\theta _{j_{1}...j_{k}}$ (or $\varphi _{j_{1}...j_{k}}$)
denote the $k$-th order partial derivatives of the dependent variables
\[
\theta _{\left( J\right) }=\theta _{j_{1}...j_{k}}=\frac{\partial ^{k}\theta
}{\partial x^{j_{1}}\partial x^{j_{2}}...\partial x^{j_{k}}}.
\]
The $J$ denotes a multi-index $J=\left( j_{1},j_{2},\ldots ,j_{k}\right) $
with $j_{k}=1,\ldots ,3$ ($k\geq 0$) pointing to the independent variables
with respect to which one differentiate.

Consider a local one-parameter Lie group of point transformations
acting on a space $\Omega $ of the independent and dependent
variables involved in the basic equations (\ref{eq:ferroeq}). The
infinitesimal generator of such a group is a vector field
$\hat{X}$ on $\Omega $
\begin{equation}
\hat{X}=\xi ^{k}\frac{\partial }{\partial x^{k}}+\eta ^{1}\frac{\partial }{%
\partial \theta }+\eta ^{2}\frac{\partial }{\partial \varphi }
\label{eq:LieBack}
\end{equation}
whose components $\;\xi ^{k}$ and $\eta ^{\alpha }$ are supposed
to be functions of class ${\mathrm{C}}^{\infty }$ on $\Omega $.
The infinitesimal invariance criterion of the equations
$\mathrm{F}=\left(\mathrm{F}^{1}, \mathrm{F}^{2}\right) $ under
the group $\mathrm{G}$ is given by the determining equations
\begin{equation}
\left( {\hat{Y}} \mathrm{F} \right) _{\mathrm{F}=0}=0,  \label{eq:determ}
\end{equation}
\begin{equation}
{\hat{Y}}=\hat{D}_{\left( J\right) }\left( W^{\alpha }\right) \partial
_{u_{\left( J\right) }^{\alpha }},  \label{eq:prolong}
\end{equation}
where $\hat{D}_{k}$ means total differentiation
\begin{equation}
\hat{D}_{k}=\partial _{k}+u_{k\left( J\right) }^{\alpha }\partial
_{u_{\left( J\right) }^{\alpha }}  \label{eq:totalD}
\end{equation}
and $W^{\alpha }=\eta ^{\alpha }-\xi ^{i}u_{i}^{\alpha }$ are the
characteristics of the canonical vector field ${\hat{Y}}$ which is
equivalent to $\hat{X}$
\begin{equation}
{\hat{Y}=}\hat{X}-\xi ^{k}\hat{D}_{k}.  \label{eq:Canon}
\end{equation}

The determining equations (\ref{eq:determ}) are realized as an over-determined
system of linear homogeneous partial differential equations with respect to
the unknown functions $\xi ^{k}$ and $\eta ^{\alpha }$. In the determining
equations the $u_{\left( J\right) }^{\alpha }$ up to second order
derivatives are the independent variables and these equations must be
fulfilled identically over them on the manifold $\mathrm{F}$.

A sufficient condition of an existence of conservation laws is an invariance
of the elementary action associated with the Euler equations under a
symmetry group of these equations. The starting point is the so-called
Noether identity \cite{ibr}
\begin{equation}
\hat{Y}+\hat{D}_{i}\cdot \xi ^{i}=W^{\alpha }\hat{E}^{\alpha }+\hat{D}_{i}%
\hat{N}^{i}.  \label{eq:Ntident}
\end{equation}
In Eq.(\ref{eq:Ntident}) $\hat{N}^{i}$ are the Noether operators given by the
expressions
\begin{equation}
\hat{N}^{i}=\xi ^{i}+\hat{D}_{i \left( s\right) }\left( W^{\alpha
}\right) \left( -1\right) ^{r}\hat{D}_{\left( r\right) }\partial
_{u_{\left( s\right) \left( r\right) }^{\alpha }}
\label{eq:Noetoper}
\end{equation}
and
\begin{equation}
\hat{E}^{\alpha }=\left( -1\right) ^{s}\hat{D}_{\left( s\right) }\partial
_{u_{\left( s\right) }^{\alpha }}  \label{eq:Euler}
\end{equation}
are the Euler-Lagrange operators. The variational symmetries on the
Euler-Lagrange equations may be found via the Noether identity

\begin{equation}
\hat{Y}(L)+\hat{D}_{i}\cdot (\xi ^{i}L)=0  \label{eq:Noether}
\end{equation}
with the Lagrangian density $L$ \cite{bostrem}
\begin{equation}
L=L(\theta ,\varphi _{t},\theta _{k},\varphi _{k})=S\left( \cos \theta
-1\right) \varphi _{t}-\beta \left[ \frac{1}{2}\sin ^{2}\theta \left( \vec{%
\nabla}\varphi \right) ^{2}+\frac{1}{2}\left( \vec{\nabla}\theta \right)
^{2}\right] .  \label{eq:Lagrange}
\end{equation}
The dot in Eq.(\ref{eq:Ntident}) means the differentiation rule $\hat{D}%
_{i}\cdot \xi ^{i}\left( L\right) \equiv L\hat{D}_{i}\left( \xi ^{i}\right)
+\xi ^{i}\hat{D}_{i}\left( L\right) $.

Substituting (\ref{eq:Lagrange}) into (\ref{eq:Noether}) one obtain
\begin{equation}
\hat{Y}(L)=W^{1}\partial _{_{\theta }}L+\hat{D}_{t}(W^{2})\partial _{\varphi
_{t}}L+\hat{D}_{k}(W^{1})\partial _{\theta _{k}}L+\hat{D}_{k}(W^{2})\partial
_{\varphi _{k}}L,  \label{eq:YL}
\end{equation}
where
\[
\partial _{_{\theta }}L=-S\sin \theta \varphi _{t}-\frac{\beta }{2}\sin
2\theta \left( \vec{\bigtriangledown}\varphi \right) ^{2},\;\partial
_{\varphi _{t}}L=S\left( \cos \theta -1\right) ,
\]
\begin{equation}
\partial _{\theta _{k}}L=-\beta \theta _{k},\;\partial _{\varphi
_{k}}L=-\beta \sin ^{2}\theta \varphi _{k}.  \label{eq:Lderivatives}
\end{equation}
After a little manipulation one get the following equation for the unknowns $%
\xi ^{k}$ and $\eta ^{\alpha }$%
\begin{eqnarray*}
&&\eta ^{1}\left( -S\sin \theta \varphi _{t}-\frac{\beta }{2}\sin 2\theta
\left( \vec{\bigtriangledown}\varphi \right) ^{2}\right) \\
&&+S\left( \cos \theta -1\right) \left( \hat{D}_{t}\left( \eta ^{2}\right)
-\varphi _{t}\hat{D}_{t}\left( \xi ^{t}\right) -\varphi _{k}\hat{D}%
_{t}\left( \xi ^{k}\right) \right) \\
&&-\beta \theta _{k}\left( \hat{D}_{k}\left( \eta ^{1}\right) -\theta _{t}%
\hat{D}_{k}\left( \xi ^{t}\right) -\theta _{l}\hat{D}_{k}\left( \xi
^{l}\right) \right) \\
&&-\beta \sin ^{2}\theta \varphi _{k}\left( \hat{D}_{k}\left( \eta
^{2}\right) -\varphi _{t}\hat{D}_{k}\left( \xi ^{t}\right) -\varphi _{l}\hat{%
D}_{k}\left( \xi ^{l}\right) \right)
\end{eqnarray*}
\begin{equation}
+\left( S\left( \cos \theta -1\right) \varphi _{t}-\frac{\beta }{2}\sin
^{2}\theta \left( \vec{\bigtriangledown}\varphi \right) ^{2}-\frac{\beta }{2}%
\left( \vec{\bigtriangledown}\theta \right) ^{2}\right) \left( \hat{D}%
_{t}\left( \xi ^{t}\right) +\hat{D}_{k}\left( \xi ^{k}\right) \right) =0.
\label{eq:Main}
\end{equation}
Hereinafter, we keep the upper index $k$ just for the space coordinates, the
corresponding time component will be written explicitly. The coefficients at
the derivatives of the functions $\theta $ and $\varphi $ constitute an
over-determined system of linear homogeneous partial differential equations
with respect to the unknowns $\xi ^{i}$ and $\eta ^{\alpha }.$ For example,
extracting the terms at the independent variable $\theta _{t}$ and set equal
them to zero
\[
S\left( \eta _{\theta }^{2}-\varphi _{k}\xi _{\theta }^{k}\right) \left(
\cos \theta -1\right) +\beta \theta _{k}\left( \xi _{k}^{t}+\theta _{k}\xi
_{\theta }^{t}+\varphi _{k}\xi _{\varphi }^{t}\right)
\]
\begin{equation}
+\beta \left( -\frac{1}{2}\sin ^{2}\theta \left( \vec{\nabla}\varphi \right)
^{2}-\frac{1}{2}\left( \vec{\nabla}\theta \right) ^{2}\right) \xi _{\theta
}^{t}=0\;  \label{eq:zerdet}
\end{equation}
one obtain the conditions
\begin{equation}
\eta _{\theta }^{2}=\xi _{\theta }^{k}=\xi _{k}^{t}=\xi _{\varphi }^{t}=\xi
_{\theta }^{t}=0  \label{eq:firstset}
\end{equation}
that immediately imply $\xi ^{t}=\xi ^{t}(t).$ The results of the full
analysis of the Eq. (\ref{eq:Main}) are summarized in Table 1.
\begin{table}
\caption{Equations for the uknowns $\xi^{k}$ and $\eta^{\alpha}$.}
\fbox{$
\begin{array}{c|c}
\begin{array}{c}
\text{Independent} \\
\text{variable}
\end{array}
& \text{Equation} \\ \hline
0 & \eta _{t}^{2}=0 \\
\varphi _{t} & \left( \cos \theta -1\right) \left( \eta _{\varphi }^{2}+div%
\vec{\xi}\right) =\eta ^{1}\sin \theta \\
\varphi _{k\text{ }} & \;-S\left( \cos \theta -1\right) \xi _{t}^{k}-\beta
\sin ^{2}\theta \eta _{k}^{2}=0\Longrightarrow \;\xi _{t}^{k}=\eta _{k}^{2}=0
\\
\theta _{k\text{ }} & \eta _{k}^{1}=0 \\
\left( \vec{\nabla}\varphi \right) ^{2} & \xi _{k}^{k}=\eta ^{1}\cot \theta +%
\frac{1}{2}\left( \xi _{t}^{t}+div\vec{\xi}\right) +\eta _{\varphi }^{2} \\
\varphi _{k}\varphi _{l}(k\neq l) & \sin ^{2}\theta \left( \xi _{k}^{l}+\xi
_{l}^{k}\right) =0\Longrightarrow \xi _{k}^{l}+\xi _{l}^{k}=0 \\
\left( \vec{\nabla}\theta \right) ^{2} & \xi _{k}^{k}=\eta _{\theta }^{1}+%
\frac{1}{2}\left( \xi _{t}^{t}+div\vec{\xi}\right) \\
\theta _{k}\theta _{l}(k\neq l) & \xi _{k}^{l}+\xi _{l}^{k}=0 \\
\theta _{k}\varphi _{k} & \eta _{\varphi }^{1}=0 \\
\left( \vec{\nabla}\varphi \right) ^{2}\varphi _{k} & \xi _{\varphi }^{k}=0
\end{array}
$}
\end{table}
Finally, we get $\xi ^{t}=\xi ^{t}(t)$, $\xi ^{k}=\xi ^{k}(\vec{r})$, $\eta
^{1}=\eta ^{1}(t,\theta )$, $\eta ^{2}=\eta ^{2}(\varphi )$. The uknowns $%
\xi ^{k}$ satisfy to Killing's equation $\xi _{k}^{l}+\xi _{l}^{k}=\sigma
\delta _{lk}$\smallskip . As is well known, it has a solution for an
arbitrary $\sigma $ just for $n\geq 3$. However, for the system considered
there is a solution in $2D$ case. Indeed, from the Table 1 one have
\[
\begin{array}{c}
\sigma =2\xi _{k}^{k}=2\eta _{\theta }^{1}+\xi _{t}^{t}+div\vec{\xi}, \\
\sigma =2\xi _{k}^{k}=2\eta ^{1}\cot \theta +\xi _{t}^{t}+div\vec{\xi}+2\eta
_{\varphi }^{2},
\end{array}
\]
(no summation over $k$) that imply
\[
\xi ^{t}=\xi ^{t}(t),\;\vec{\xi}(\vec{r})=\vec{a}+\hat{B}\vec{r},\;\hat{B}%
^{T}=-\hat{B}=-\left(
\begin{array}{cc}
0 & -b \\
b & 0
\end{array}
\right) ,
\]
\[
{\eta}^{1}=0,\; {\eta}^{2}={\upsilon} .
\]
The solution found results in the following expressions for the
characteristics $W^{1}=-\xi ^{t}\theta _{t}-\vec{a}\vec{\bigtriangledown}%
\theta -b\left[ \vec{r}\times \vec{\bigtriangledown}\theta \right] _{z}$ and
$W^{2}=\upsilon -\xi ^{t}\varphi _{t}-\vec{a}\vec{\bigtriangledown}\varphi
-b\left[ \vec{r}\times \vec{\bigtriangledown}\varphi \right] _{z}$. One can
derive corresponding conservation laws through Noether's theorem procedure
\begin{equation}
\hat{D}_{i}\hat{N}^{i}\left( L\right) =0  \label{eq:current}
\end{equation}
under the condition $\hat{E}^{\alpha }L=0.$ The Noether operators modify the
Lagrangian density into a conserved quantity $C^{i}=\hat{N}^{i}\left(
L\right) $ with a zero total divergence. Using the found variational
symmetries we first construct the operators $\hat{N}^{i}$ from formulae (\ref{eq:Noetoper}) and then calculate from (\ref{eq:current}) the corresponding conservation laws. The results are brought together in the Table 2.
\begin{table}
\caption{Conservation laws.}
\frame{$
\begin{array}{c|c}
& \text{Conservation} \\
& \text{law} \\ \hline
1 & \hat{D}_{t}\left( \frac{\beta }{2}\sin ^{2}\theta \left( \vec{\nabla}%
\varphi \right) ^{2}+\frac{\beta }{2}\left( \vec{\nabla}\theta \right)
^{2}\right) =\hat{D}_{k}\left( \beta \theta _{t}\theta _{k}+\beta \varphi
_{t}\varphi _{k}\sin ^{2}\theta \right) \\
2 & \hat{D}_{t}\left( S\left( \cos \theta -1\right) \varphi _{k}\right) =%
\hat{D}_{l}\left( T_{kl}\right) \\
3 & \hat{D}_{t}\left( \left[ \vec{r}\times S\left( \cos \theta -1\right)
\vec{\nabla}\varphi \right] _{l}\right) =\hat{D}_{k}\left( \epsilon
^{lmn}x^{m}T_{nk}\right) ,\;\left( l=z\right) \\
4 & \hat{D}_{t}\left( S\left( \cos \theta -1\right) \right) =\hat{D}%
_{k}\left( \beta \varphi _{k}\sin ^{2}\theta \right)
\end{array}
$}
\end{table}
The physical interpretation of the obtained relations is obvious.

1) The symmetry group of translations in time $t\rightarrow t+a$ yields to
the energy conservation law with the energy density $W=\frac{JS^{2}}{2}\sin
^{2}\theta \left( \vec{\nabla}\varphi \right) ^{2}+\frac{JS^{2}}{2}\left(
\vec{\nabla}\theta \right) ^{2}$ and the energy density current $\vec{T}%
=-(JS^{2}\theta _{t}\vec{\nabla}\theta +JS^{2}\varphi _{t}\vec{\nabla}%
\varphi \sin ^{2}\theta )$.

2) This law connects the momentum $\vec{P}=\hbar S\left( 1-\cos \theta
\right) \vec{\nabla}\varphi $ and the canonical energy-momentum tensor
\begin{equation}
T_{kl}=\hbar L\delta _{kl}+\hbar \beta \theta _{k}\theta _{l}+\hbar \beta
\varphi _{k}\varphi _{l}\sin ^{2}\theta  \label{eq:tensoren}
\end{equation}
and reflects the variational symmetry under space translations $%
\;x_{l}\rightarrow x_{l}+a_{l}.$

3) The conservation of angular momenta $\vec{L}=\left[ \vec{r}\times \vec{P}%
\right] $ through the angular momenta tensor under the space rotation around
$z$-axis.

4) The conservation law corresponding to the transformations $\varphi
\rightarrow \varphi +\chi $ connects the scalar variable $N=S\left( 1-\cos
\theta \right) $ with the current $\vec{j}=\beta \vec{\nabla}\varphi \sin
^{2}\theta $.

It is useful to apply the obtained laws to some types of solutions of Eqs. (\ref{eq:ferroeq}). We restrict consideration by finite-amplitude spin-waves, dynamical soliton and skyrmion.

1) The finite-amplitude spin-waves $\theta =\theta _{0}=const$ and $\varphi =%
\vec{k}\vec{r}-\omega t$ \cite{PhysRep} with the dispersion $\hbar \omega =JS%
\vec{k}^{2}\cos \theta _{0}$. The last conservation law may be interpreted
as a conservation of the magnon density $N=S\left( 1-\cos \theta _{0}\right)
$. The corresponding magnon density current is $\vec{j}=\frac{JS^{2}}{\hbar }%
\vec{k}\sin ^{2}\theta _{0}$. The space components of the energy-momentum
tensor are
\[
\hat{T}_{kl}=\left[
\begin{array}{cc}
\begin{array}{c}
\begin{array}{c}
S\left( 1-\cos \theta _{0}\right) \hbar \omega \\
+\frac{JS^{2}}{2}\sin ^{2}\theta _{0}\left( k_{x}^{2}-k_{y}^{2}\right)
\end{array}
\end{array}
&
\begin{array}{c}
JS^{2}\sin ^{2}\theta _{0}k_{x}k_{y}
\end{array}
\\
JS^{2}\sin ^{2}\theta _{0}k_{x}k_{y} &
\begin{array}{c}
S\left( 1-\cos \theta _{0}\right) \hbar \omega \\
-\frac{JS^{2}}{2}\sin ^{2}\theta _{0}\left( k_{x}^{2}-k_{y}^{2}\right)
\end{array}
\end{array}
\right] .
\]

The momentum $\vec{P}=S\left( 1-\cos \theta _{0}\right) \hbar \vec{k}$ is a
times of the magnon density and the elementary momentum $\hbar \vec{k}$. The
energy density $W=\frac{1}{2}JS^{2}k^{2}\sin ^{2}\theta _{0}\approx N\hbar
\omega $ ($\theta _{0}\ll 1$) has the corresponding energy density current $%
\vec{T}=JS^{2}\vec{k}\omega \sin ^{2}\theta _{0}$. We note a non-additive
character of energy density at finite values $\theta _{0}$.

2) For a dynamical soliton with axial symmetry one have to use the following
parametrization $\theta =\theta (r)$ and $\varphi =\omega t$ \cite{PhysRep}.
The magnon density $N=S\left( 1-\cos \theta (r)\right) $ has the zero magnon
density current $\vec{j}=0$. The energy-momentum tensor is
\[
\hat{T}_{kl}=\left[
\begin{array}{cc}
\begin{array}{c}
\begin{array}{c}
-S\left( 1-\cos \theta \left( r\right) \right) \hbar \omega \\
+\frac{JS^{2}}{2}\cos \left( 2\omega t\right) \left( \frac{d\theta }{dr}%
\right) ^{2}
\end{array}
\end{array}
&
\begin{array}{c}
\frac{JS^{2}}{2}\sin \left( 2\omega t\right) \left( \frac{d\theta }{dr}%
\right) ^{2}
\end{array}
\\
\frac{JS^{2}}{2}\sin \left( 2\omega t\right) \left( \frac{d\theta }{dr}%
\right) ^{2} &
\begin{array}{c}
-S\left( 1-\cos \theta \left( r\right) \right) \hbar \omega \\
-\frac{JS^{2}}{2}\cos \left( 2\omega t\right) \left( \frac{d\theta }{dr}%
\right) ^{2}
\end{array}
\end{array}
\right] .
\]
The momentum $\vec{P}$ is zero, the energy density $W=\frac{JS^{2}}{2}\left(
\frac{d\theta }{dr}\right) ^{2}$ and the energy density current $\vec{T}=0$.

3) For the skyrmion the following parametrization $\theta =2\tan ^{-1}\left(
\frac{R}{r}\right) $ and $\varphi =\tan ^{-1}\left( \frac{y}{x}\right) $
\cite{Belavin} is used (for simplicity we consider an unit topological
charge). Then, $N=2S\frac{R^{2}}{r^{2}+R^{2}}$, $j_{r}=0$, $j_{\varphi }=%
\frac{JS^{2}}{\hbar }\frac{4R^{2}r}{\left( R^{2}+r^{2}\right) ^{2}}$. The
momentum in polar coordinates of $xy$-plane $P_{r}=0$, $P_{\varphi }=\frac{%
2\hbar S}{r}\frac{R^{2}}{r^{2}+R^{2}}$, however, $\hat{T}_{kl}=0$. The
energy density $W=4JS^{2}\frac{R^{2}}{\left( R^{2}+r^{2}\right) ^{2}}=\frac{%
JL_{z}^{2}}{\hbar ^{2}R^{2}}$ and the energy density current $\vec{T}=0$.

\section{Dynamical invariants of $2D$ two-sublattice magnet.}

In this section Lie transformation group methods will be applied to the
partial differential equations describing the dynamics of $2D$ generic
antiferromagnet (two-sublattice magnet). The established variational Lie
symmetries for ferromagnet will be employed to derive antiferromagnet
conservation laws revealing its important features.

The action of $2D$ two-sublattice magnet is \cite{bostrem}
\begin{eqnarray}
L &=&S\sum\limits_{i=1}^{2}\left( \cos \theta _{i}-1\right) \varphi
_{it}-\beta \left( -2\sin \theta _{1}\sin \theta _{2}\cos \left( \varphi
_{1}-\varphi _{2}\right) -2\cos \theta _{1}\cos \theta _{2}\right.  \nonumber
\\
&&+\cos \theta _{1}\cos \theta _{2}\cos \left( \varphi _{1}-\varphi
_{2}\right) \left( \vec{\nabla}\theta _{1}\vec{\nabla}\theta _{2}\right)
-\sin \theta _{1}\cos \theta _{2}\sin \left( \varphi _{1}-\varphi
_{2}\right) \left( \vec{\nabla}\varphi _{1}\vec{\nabla}\theta _{2}\right)
\nonumber \\
&&+\cos \theta _{1}\sin \theta _{2}\sin \left( \varphi _{1}-\varphi
_{2}\right) \left( \vec{\nabla}\theta _{1}\vec{\nabla}\varphi _{2}\right)
+\sin \theta _{1}\sin \theta _{2}\cos \left( \varphi _{1}-\varphi
_{2}\right) \left( \vec{\nabla}\varphi _{1}\vec{\nabla}\varphi _{2}\right)
\nonumber \\
&&\left. +\sin \theta _{1}\sin \theta _{2}\left( \vec{\nabla}\theta _{1}\vec{%
\nabla}\theta _{2}\right) \right) .  \label{eq:LagrAFM}
\end{eqnarray}
The time symmetry group $\;t\rightarrow t+a$ under which the elementary
action is invariant has the infinitesimal symmetry generator $\hat{X}=\frac{%
\partial }{\partial t}$ and Noether's operators $\hat{N}^{t}=1-\sum%
\limits_{i=1}^{2}\theta _{it}\frac{\partial }{\partial \theta _{it}}%
-\sum\limits_{i=1}^{2}\varphi _{it}\frac{\partial }{\partial \varphi _{it}}$%
, $\hat{N}^{k}=-\sum\limits_{i=1}^{2}\theta _{it}\frac{\partial }{\partial
\theta ik}-\sum\limits_{i=1}^{2}\varphi _{it}\frac{\partial }{\partial
\varphi _{ik}}$. By introducing the functions
\begin{eqnarray}
C^{t} &=&\hat{N}^{t}\left( L\right) =-\beta \left( -2\sin \theta _{1}\sin
\theta _{2}\cos \left( \varphi _{1}-\varphi _{2}\right) -2\cos \theta
_{1}\cos \theta _{2}\right.  \nonumber \\
&&+\cos \theta _{1}\cos \theta _{2}\cos \left( \varphi _{1}-\varphi
_{2}\right) \left( \vec{\nabla}\theta _{1}\vec{\nabla}\theta _{2}\right)
-\sin \theta _{1}\cos \theta _{2}\sin \left( \varphi _{1}-\varphi
_{2}\right) \left( \vec{\nabla}\varphi _{1}\vec{\nabla}\theta _{2}\right)
\nonumber \\
&&+\cos \theta _{1}\sin \theta _{2}\sin \left( \varphi _{1}-\varphi
_{2}\right) \left( \vec{\nabla}\theta _{1}\vec{\nabla}\varphi _{2}\right)
+\sin \theta _{1}\sin \theta _{2}\cos \left( \varphi _{1}-\varphi
_{2}\right) \left( \vec{\nabla}\varphi _{1}\vec{\nabla}\varphi _{2}\right)
\nonumber \\
&&\left. +\sin \theta _{1}\sin \theta _{2}\left( \vec{\nabla}\theta _{1}\vec{%
\nabla}\theta _{2}\right) \right)  \label{eq:Cten}
\end{eqnarray}

and
\begin{eqnarray}
C^{k} &=&\hat{N}^{k}\left( L\right) =\beta \left( \cos \theta _{1}\cos
\theta _{2}\cos \left( \varphi _{1}-\varphi _{2}\right) +\sin \theta
_{1}\sin \theta _{2}\right) \left( \theta _{1t}\theta _{2k}+\theta
_{2t}\theta _{1k}\right)  \nonumber \\
&&+\beta \sin \theta _{1}\sin \theta _{2}\cos \left( \varphi _{1}-\varphi
_{2}\right) \left( \varphi _{1t}\varphi _{2k}+\varphi _{2t}\varphi
_{1k}\right)  \nonumber \\
&&+\beta \cos \theta _{1}\sin \theta _{2}\sin \left( \varphi _{1}-\varphi
_{2}\right) \left( \theta _{1t}\varphi _{2k}+\theta _{1k}\varphi _{2t}\right)
\nonumber \\
&&-\beta \sin \theta _{1}\cos \theta _{2}\sin \left( \varphi _{1}-\varphi
_{2}\right) \left( \theta _{2t}\varphi _{1k}+\theta _{2k}\varphi _{1t}\right)
\label{eq:Cken}
\end{eqnarray}
one get the energy conservation law $\hat{D}_{t}\left( C^{t}\right) +\hat{D}%
_{k}\left( C^{k}\right) =0$. By the same manner one can obtain the
generators and Noether's operators of the translation group $%
x_{k}\rightarrow x_{k}+a_{k}$

$\hat{X}=\frac{\partial }{\partial x^{k}}$, $\hat{N}^{t}=-\sum%
\limits_{i=1}^{2}\theta _{ik}\frac{\partial }{\partial \theta it}%
-\sum\limits_{i=1}^{2}\varphi _{ik}\frac{\partial }{\partial \varphi _{it}}$%
, $\hat{N}^{l}=\delta _{kl}-\sum\limits_{i=1}^{2}\theta _{jk}\frac{\partial
}{\partial \theta _{jl}}-\sum\limits_{i=1}^{2}\varphi _{jk}\frac{\partial }{%
\partial \varphi _{jl}}$.

The momentum conservation law is
\begin{equation}
\sum\limits_{i=1}^{2}\hat{D}_{t}\left( S\left( \cos \theta _{i}-1\right)
\varphi _{ik}\right) =\hat{D}_{l}\left( T_{kl}\right) ,  \label{eq:momAFM}
\end{equation}
with the tensor
\begin{equation}
T_{kl}=L\delta _{kl}-\sum\limits_{j=1}^{2}\theta _{jk}\partial _{\theta
_{jk}}L-\sum\limits_{j=1}^{2}\varphi _{jk}\partial _{\varphi _{jk}}L.
\label{eq:tensenimp}
\end{equation}

The conservation of angular momenta is
\begin{equation}
\sum\limits_{j=1}^{2}\hat{D}_{t}\left( \left[ \vec{r}\times S\left( \cos
\theta _{j}-1\right) \vec{\nabla}\varphi _{j}\right] _{z}\right) =\hat{D}%
_{k}\left( \epsilon ^{znl}x^{n}T_{lk}\right) ,  \label{eq:TENIMPAFM}
\end{equation}
\[
\hat{X}=\epsilon ^{znl}x^{n}\frac{\partial }{\partial x^{l}},
\]
\[
\hat{N}^{t}=-\epsilon ^{znl}x^{n}\theta _{l}^{\alpha }\frac{\partial }{%
\partial \theta _{t}^{\alpha }}-\epsilon ^{znl}x^{n}\varphi _{l}^{\alpha }%
\frac{\partial }{\partial \varphi _{t}^{\alpha }},
\]
\[
\hat{N}^{k}=\epsilon
^{znk}x^{n}-\epsilon ^{znl}x^{n}\theta _{l}^{\alpha }\frac{\partial }{%
\partial \theta _{k}^{\alpha }}-\epsilon ^{znl}x^{n}\varphi _{l}^{\alpha }%
\frac{\partial }{\partial \varphi _{k}^{\alpha }}.
\]
The action is also invariant under a simultaneous rotation in both
sublattices $\varphi _{1}^{^{\prime }}=\varphi _{1}+\chi $, $\varphi
_{2}^{^{\prime }}=\varphi _{2}+\chi $. The generator and Noether operators
are $\hat{X}=\sum\limits_{j=1}^{2}\frac{\partial }{\partial \varphi _{j}}$, $%
\hat{N}^{t}=\sum\limits_{j=1}^{2}\frac{\partial }{\partial \varphi _{jt}}$, $%
\hat{N}^{k}=\sum\limits_{j=1}^{2}\frac{\partial }{\partial \varphi _{jk}}$
and they imply the conservation law
\begin{eqnarray}
\sum\limits_{j=1}^{2}\hat{D}_{t}\left( S\cos \theta _{j}\right) &=&\beta
\hat{D}_{k}\left( \sin \theta _{1}\sin \theta _{2}\cos \left( \varphi
_{1}-\varphi _{2}\right) \left( \varphi _{1k}+\varphi _{2k}\right) \right.
\nonumber \\
&&\left. -\sin \left( \varphi _{1}-\varphi _{2}\right) \left( \theta
_{2k}\sin \theta _{1}\cos \theta _{2}-\theta _{1k}\sin \theta _{2}\cos
\theta _{1}\right) \right) .  \label{eq:NUMAFM}
\end{eqnarray}

\section{\protect\smallskip Conclusions.}

Lie transformation group methods have been applied to the partial
differential equations describing the dynamics of ferro and antiferromagnets
and the variational symmetries of the action functional on the
Euler-Lagrange equations are established. Each such symmetry gives rise to a
conservation law with a conserved current. The explicit expressions for the
conserved currents are found.

\smallskip

\vspace{1cm} {\Large Appendix}.

Let's take the following notations:

1) the matrix of Frechet derivatives $\hat{f}_{\beta }^{\alpha
}=\sum\limits_{s}\frac{\partial f^{\alpha }}{\partial u_{\left( s\right)
}^{\beta }}\hat{D}_{\left( s\right) }$, where the sum is taken over all
multi-indices $s=\left( s_{1},s_{2},\ldots ,s_{k}\right) $ with $%
s_{k}=1,\ldots ,3$ ($k\geq 0$);

2) the adjoint operator of Frechet derivatives \cite{Olver}
\[
\left( \hat{f}_{\alpha }^{\beta }\right) ^{*}=\sum\limits_{s}\left( -\hat{D}%
\right) _{\left( s\right) }\cdot \frac{\partial f^{\alpha }}{\partial
u_{\left( s\right) }^{\beta }}.
\]
One have to found a Lagrangian density $L\left( x,u,u^{^{\prime }},\ldots
\right) $ that the initial set of differential equations are the
Euler-Lagrange equations of the following action
\[
S[u]=\int L\left( x,u,u^{^{\prime }},\ldots \right) {d x} \equiv
\int\limits_{t_{2}}^{t_{1}}{d t}\int {d V} L\left( x,u,u^{^{\prime
}},\ldots \right) .
\]

From the ratio
\[
\hat{E}^{\alpha }L=f^{\alpha }+\int\limits_{0}^{1}\left[ \hat{f}^{\dagger }-%
\hat{f}\right] _{\beta }^{\alpha }\left[ \lambda u^{\beta }\left(
x\right) \right] {d \lambda}
\]
is seen that that sufficient condition of the solution of the problem is the
self-adjoint condition of the $\hat{f}$-operator
\[
\frac{\partial f^{\alpha }}{\partial u_{\left( s\right) }^{\beta }}=\left(
-1\right) ^{s+r}C_{s+r}^{s}\hat{D}_{\left( r\right) }\frac{\partial f^{\beta
}}{\partial u_{\left( s\right) \left( r\right) }^{\alpha }}
\]
or in the matrix notations $\hat{f}_{\beta }^{\alpha }=\left( \hat{f}%
^{\dagger}\right) _{\beta }^{\alpha }=\left( \hat{f}_{\alpha }^{\beta
}\right) ^{*}$. The straight calculation yields to

\[
\hat{f}=\left(
\begin{array}{cc}
-S\varphi _{t}\cos \theta -\beta \left( \vec{\nabla}\varphi \right) ^{2}\cos
2\theta +\beta \Delta & -S\sin \theta \partial _{t}-\beta \sin 2\theta
\left( \vec{\nabla}\varphi \vec{\nabla}\right) \\
\begin{array}{c}
S\theta _{t}\cos \theta +2\beta \cos 2\theta \left( \vec{\nabla}\varphi \vec{%
\nabla}\theta \right) +\beta \sin 2\theta \Delta \varphi \\
+S\sin \theta \partial _{t}+\beta \sin 2\theta \left( \vec{\nabla}\varphi
\vec{\nabla}\right)
\end{array}
& \beta \sin 2\theta \left( \vec{\nabla}\varphi \vec{\nabla}\right) +\beta
\sin ^{2}\theta \Delta
\end{array}
\right) ,
\]
and

\[
\hat{f}^{*}=\left(
\begin{array}{cc}
-S\varphi _{t}\cos \theta -\beta \left( \vec{\nabla}\varphi \right) ^{2}\cos
2\theta +\beta \Delta &
\begin{array}{c}
S\theta _{t}\cos \theta +2\beta \cos 2\theta \left( \vec{\nabla}\varphi \vec{%
\nabla}\theta \right) +\beta \sin 2\theta \Delta \varphi \\
+S\sin \theta \partial _{t}+\beta \sin 2\theta \left( \vec{\nabla}\varphi
\vec{\nabla}\right)
\end{array}
\\
-S\sin \theta \partial _{t}-\beta \sin 2\theta \left( \vec{\nabla}\varphi
\vec{\nabla}\right) & \beta \sin 2\theta \left( \vec{\nabla}\varphi \vec{%
\nabla}\right) +\beta \sin ^{2}\theta \Delta
\end{array}
\right) .
\]
Obviously, the claimed condition is fulfilled. The Lagrangian density can be
constructed via the homotopy formula \cite{Olver}
\[
L[u]=\sum\limits_{\alpha }u^{\alpha }\left( x\right)
\int\limits_{0}^{1}f^{\alpha }\left( x,\lambda u,\lambda
u^{^{\prime }},\ldots \right) {d \lambda}.
\]
Using the explicit form of differential equations it results in
\begin{eqnarray*}
L &=&-S\left( \theta \varphi _{t}-\varphi \theta _{t}\right)
\int\limits_{0}^{1}{d \lambda}\;\lambda \sin \lambda \theta +\beta
\theta
\Delta \theta \int\limits_{0}^{1}{d \lambda} \;\lambda \\
&&+\beta \left( \varphi \left( \vec{\nabla}\varphi
\vec{\nabla}\theta \right) -\frac{\theta }{2}\left(
\vec{\nabla}\varphi \right) ^{2}\right) \int\limits_{0}^{1}{d
\lambda} \;\lambda ^{2}\sin 2\lambda \theta +\beta \varphi \Delta
\varphi \int\limits_{0}^{1}{d \lambda} \;\lambda \sin ^{2}\lambda
\theta ,
\end{eqnarray*}

or after some manipulations in
\begin{eqnarray*}
L &=&S\left( \cos \theta -1\right) \varphi _{t}-\frac{\beta }{2}\sin
^{2}\theta \left( \vec{\nabla}\varphi \right) ^{2}-\frac{\beta }{2}\left(
\vec{\nabla}\theta \right) ^{2} \\
&&+S\hat{D}_{t}\left[ \left( 1-\frac{\sin \theta }{\theta }\right) \varphi
\right] +\hat{D}_{k}\left[ \frac{\beta }{2}\theta \vec{\nabla}\theta +\frac{%
\beta }{\theta ^{2}}C_{0}\varphi \vec{\nabla}\varphi \right] ,
\end{eqnarray*}
where
\[
C_{0}=\frac{1}{4}\left[ \theta \left( \theta -\sin 2\theta \right) +\sin
^{2}\theta \right].
\]
The last two terms present total derivatives and may be dropped.

\smallskip

\acknowledgements
This work was supported by the grant NREC-005 of
US CRDF (Civilian Research \& Development Foundation).


\begin{thebibliography}{9}
\bibitem{ibr}  N.H. Ibragimov, {\em Transformation Groups applied to
Mathematical Physics \/} (Reidel, Boston, 1985).

\bibitem{Olver}  P. Olver, {\em Applications of Lie groups to differential
equations \/} (Springer-Verlag, 1986).

\bibitem{ovs}   L.V. Ovsiannikov, {\em Group Analysis of Differential
Equations \/} (Academic Press, New York, 1982).

\bibitem{bishop}  R. Balakrishnan, A.R. Bishop,
{\em Phys. Rev. B \/} {\bf 40}  (1989) 9194.

\bibitem{bostrem}  I.G. Bostrem, A.S. Ovchinnikov, R.F. Egorov, {\em Phys. Lett.
A \/}  {\bf 279} (2001) 33.

\bibitem{PhysRep}  A.M. Kosevich, B.A. Ivanov, A.S. Kovalev, {\em Phys. Rep. \/} {\bf 194}
(1990) 119.

\bibitem{Belavin}  A.A. Belavin, A.M. Polyakov, {\em JETP Letters \/} {\bf 22}  (1975) 245.
\end{thebibliography}
\end{document}